# Design Considerations for the Virtual Source/Virtual Destination (VS/VD) Feature in the ABR Service of ATM Networks

**Shiv Kalyanaraman, Raj Jain, Jianping Jiang, Rohit Goyal, Sonia Fahmy**

The Ohio State University

Department of CIS, Columbus, OH 43210-1277

Phone: +1 614 688 4482. Fax: +1 614 292 2911

Email: {*shivkuma, jain, goyal, fahmy*}@cis.ohio-state.edu

**Pradeep Samudra**

Director, System Engineering,

Broadband Network Lab, Samsung Telecom America, Inc.

1130 E Arapaho, Richardson, TX 75081.

Phone: (972) 761-7865, email: psamudra@telecom.sna.samsung.com

## Abstract

The Available Bit Rate (ABR) service in ATM networks has been specified to allow fair and efficient support of data applications over ATM utilizing capacity left over after servicing higher priority classes. One of the architectural features in the ABR specification [1] is the Virtual Source/Virtual Destination (VS/VD) option. This option allows a switch to divide an end-to-end ABR connection into separately controlled ABR segments by acting like a destination on one segment, and like a source on the other. The coupling in the VS/VD switch between the two ABR control segments is implementation specific. In this paper, we model a VS/VD ATM switch and study the issues in designing coupling between ABR segments. We identify a number of implementation options for the coupling. A good choice significantly improves the stability and transient performance of the system and reduces the buffer requirements at the switches.

## 1 Introduction

Asynchronous Transfer Mode (ATM) networks provide multiple classes of service tailored to support data, voice, and video applications. Of these, the Available Bit Rate (ABR) and the Unspecified Bit Rate (UBR) service classes have been specifically developed to support data applications. Traffic is controlled intelligently in ABR using a rate-based closed-loop end-to-end traffic management framework [1, 2, 3]. The network switches monitor available capacity and give feedback to the sources asking them to change their transmission rates. Several switch algorithms have been developed [4, 5, 6, 7, 8] to calculate feedback intelligently. The resource management (RM) cells (which carry feedback from the switches) travel from the source to the destination and back.

One of the options of the ABR framework is the Virtual Source/Virtual Destination (VS/VD) option. This option allows a switch to divide an ABR connection into separately controlled ABR segments. On one segment, the switch behaves as a destination end system, i.e., it receives data and turns around resource management (RM) cells (which carry rate feedback) to the source end system. On the other segment the switch behaves as a source end system, i.e., it controls the transmission rate of every virtual circuit (VC) and schedules the sending of data and RM cells. We call such a switch a "VS/VD switch". In effect, the end-to-end control is replaced by segment-by-segment control as shown in Figure 1.

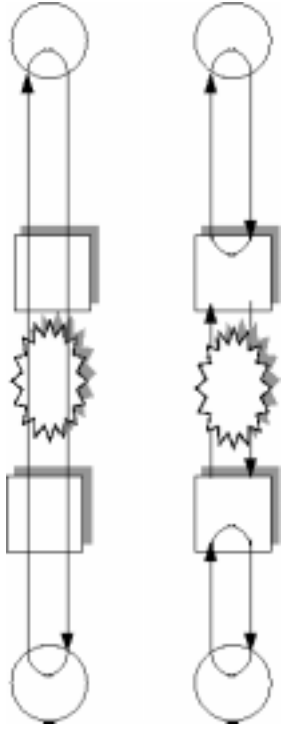

Figure 1: End-to-End Control vs VS/VD Control

One advantage of the segment-by-segment control is that it isolates different networks from each other. One example is a proprietary network like frame-relay or circuit-switched network between two ABR segments, which allows end-to-end ABR connection setup across the proprietary network and forwards ATM packets between the ABR segments[1]. Another example is the interface point between a satellite network and a LAN. The gateway switches at the edge of a satellite network can implement VS/VD to isolate downstream workgroup switches from the effects of the long delay satellite paths (like long queues).

A second advantage of segment-by-segment control is that the segments have shorter feedback loops which can potentially improve performance because feedback is given faster to the sources whenever new traffic bursts are seen.

---

[1]Signaling support for this possibility is yet to be considered by the ATM Forum

The VS/VD option requires the implementation of per-VC queueing and scheduling at the switch. In addition to per-VC queueing and scheduling, there is an incremental cost to enforce the (dynamically changing) rates of VCs, and to implement the logic for the source and destination end system rules as prescribed by the ATM Forum [1].

The goal of this study is find answers to the following questions:

- Do VS/VD switches really improve ABR performance?
- What changes to switch algorithms are required to operate in VS/VD environments?
- Are there any side-effects of having multiple control loops in series?

In this paper, we model and study VS/VD switches using the ERICA switch algorithm [8] to calculate rate feedback. We describe our switch model and the use of the ERICA algorithm in sections 2 and 3. The VS/VD design options are listed and evaluated in sections 4 and 5. The results and future work are summarized in sections 7 and 8.

## 2  Switch Queue Structure

In this section, we first present a simple switch queue model for the non-VS/VD switch and later extend it to a VS/VD switch by introducing per-VC queues. The flow of data, forward RM (FRM) and backward RM (BRM) cells is also closely examined.

### 2.1  A Non-VS/VD Switch

A minimal non-VS/VD switch has a separate FIFO queue for each of the different service classes (ABR, UBR etc.). We refer to these queues as "per-class" queues. The ABR switch rate allocation algorithm is implemented at every ABR class queue. This model of a non-VS/VD switch based network with per-class queues is illustrated in Figure 2.

Besides the switch, the figure shows a source end system, S, and a destination end system, D, each having per-VC queues to control rates of individual VCs. For example, ABR VCs control their Allowed Cell Rates (ACRs) based upon network feedback. We assume that the source/destination per-VC queues feed into corresponding per-class queues (as shown in the figure) which in turn feed to the link. This assumption is not necessary in practice, but simplifies the presentation of the model. The contention for link access between cells from different per-class queues (at the switch, the source and the destination) is resolved through appropriate scheduling.

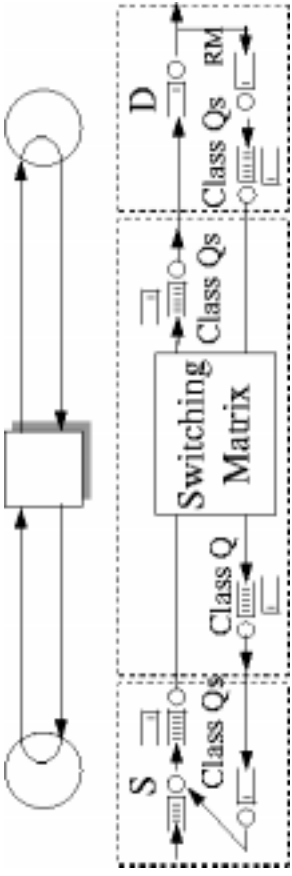

Figure 2: Per-class queues in a non-VSVD switch

## 2.2 A VS/VD Switch

The VS/VD switch implements the source and the destination end system functionality in addition to the normal switch functionality. Therefore, like any source and destination end-system, it requires per-VC queues to control the rates of individual VCs. The switch queue structure is now more similar to the source/destination structure where we have per-VC queues feeding into the per-class queues before each link. This switch queue structure and a unidirectional VC operating on it is shown in Figure 3.

The VS/VD switch has two parts. The part known as the Virtual Destination (VD) forwards the data cells from the first segment ("previous loop") to the per-VC queue at the Virtual Source (VS) of the second segment ("next loop"). The other part or the Virtual Source (of the second segment) sends out the data cells and generates FRM cells as specifed in the source end system rules.

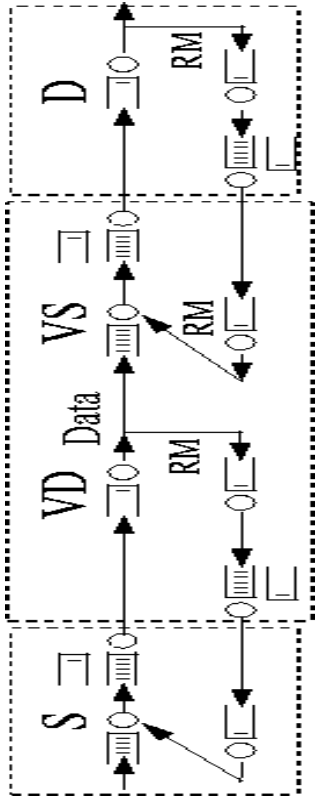

Figure 3: Per-VC and per-class queues in a VSVD switch

The switch also needs to implement the switch congestion control algorithm and calculate the allocations for VCs depending upon its bottleneck rate. A question which arises is where the rate calculations are done and how the feedback is given to the sources. We postpone the discussion of this question to later sections.

## 2.3  A VS/VD Switch with Unidirectional Data Flow

The actions of the VS/VD switch upon receiving RM cells are as follows. The VD of the previous loop turns around FRM cells as BRM cells to the VS on the same segment (as specified in the destination end system rules [2]). Additionally, when the FRM cells are turned around, the switch may decrease the value of the explicit rate (ER) field to account for the bottleneck rate of the next link and the ER from the subsequent ABR segments.

When the VS at the next loop receives a BRM cell, the ACR of the per-VC queue at the VS is updated using the ER field in the BRM (ER of the subsequent ABR segments) as specified in the source end system rules [2]). Additionally, the ER value of the subsequent ABR segments needs to be made known to the VD of the first segment. One way of doing this is for the VD of the first segment to use the ACR of the VC in the VS of the next segment while turning around FRM cells.

The model can be extended to multiple unidirectional VCs in a straightforward way. Figure 4 shows two unidirectional VCs, VC1 and VC2, between the same source S and destination D which go from Link1 to Link2 on a VS/VD switch. Observe that there is a separate VS and VD control for each VC. We omit non-ABR queues in this and subsequent figures.

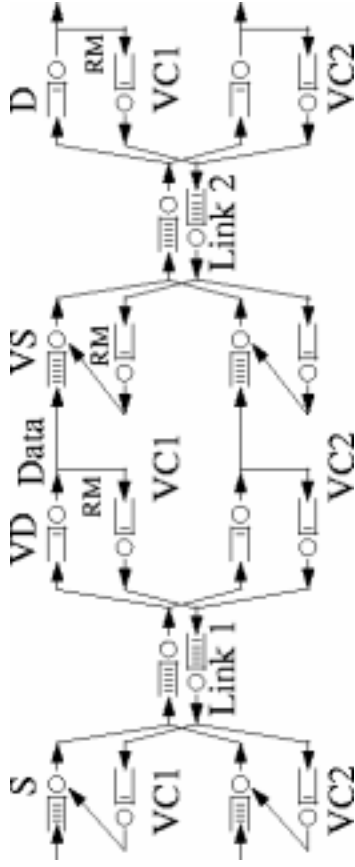

Figure 4: Multiple unidirectional VCs in a VSVD switch

## 2.4 Bi-directional Data Flow

Bi-directional flow in a VS/VD switch (Figure 5) is again a simple extension to the above model. The data on the previous loop VD is forwarded to the next loop VS. FRMs are turned around by the previous loop VD to the previous loop VS. BRMs are processed by the next loop VS to update the corresponding ACRs.

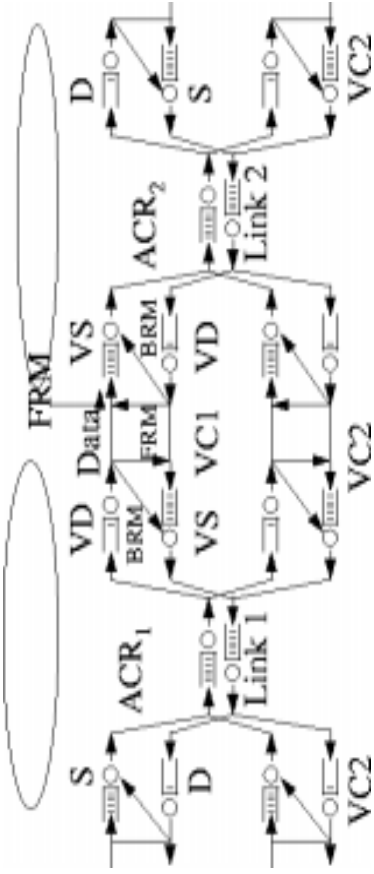

Figure 5: Multiple bi-directional VCs in a VSVD switch

We will discuss the rates and allocations of *VC1 only*. VC1 has two ACRs: $ACR_1$ in the reverse direction on Link1 and $ACR_2$ in the forward direction on Link2. Henceforth, the subscript 1 refers to the "previous loop" variables and subscript 2 to the "next loop" variables of VC1.

## 3 Basic ERICA Switch Scheme

We use the ERICA algorithm [8] for congestion control at the switches. We give a brief overview of the algorithm in this section.

ERICA first sets a target rate as follows:

Target Rate = Target Utilization × Link Rate - VBR Rate - CBR Rate

It also measures the input rate to the ABR queue and the number of active ABR sources.

To achieve fairness, the *VC*'s Allocation (VA) has a component:

$VA_{fairness}$ = Target Rate / Number of Active VCs

To achieve efficiency, the *VC*'s Allocation (VA) has a component:

$VA_{efficiency}$ = VC's Current Cell Rate / Overload, where Overload = Input Rate / Target Rate;

Finally, the *VC*'s allocation on this link (VAL) is calculated as:

VAL = Max{ VA$_{efficiency}$, VA$_{fairness}$ } = Function{ Input Rate, VC's current rate }

Note that the full ERICA algorithm contains several enhancements which account for fairness, queueing delays, and which handles highly variant bursty (ON-OFF) traffic efficiently. A complete description of the algorithm is provided in reference [8]. We now describe where the ERICA rate calculations are done in a non-VS/VD switch and in a VS/VD switch.

## 3.1 Rate Calculations in a non-VS/VD Switch

The non-VS/VD switch calculates the rate (VAL) for sources when the BRMs are processed in the reverse direction and enters it in the BRM field as follows:

ER in BRM = Min{ ER in BRM, VAL }

At the source end system, the ACR is updated as:

ACR = Function{ ER, VC's current ACR }

## 3.2 Rate Calculations in a VS/VD Switch

Figure 6 shows the rate calculations in a VS/VD switch. Specifically, the segment starting at Link2 ("next loop") returns an ER value, $ER_2$ in the BRM, and the FRM of the first segment ("previous loop") is turned around with an ER value of $ER_1$. The ERICA algorithm for the port to Link2 calculates a rate ($VAL_2$) as: $VAL_2$ = Function { Input Rate, VC's Current Rate }. The rate calculations at the VS and VD are as follows:

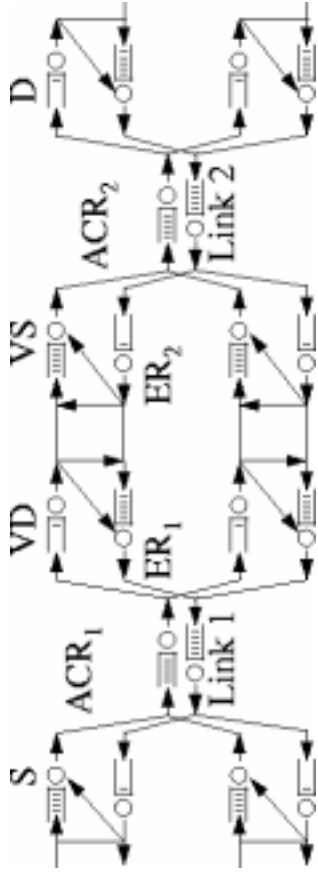

Figure 6: Rate calculations in VS/VD switches

- Destination algorithm for the *previous loop*:

  $ER_1$ = Min { $ER_1, VAL_2, ACR_2$ }

- Source Algorithm for the *next loop*:

  Optionally, $ER_2 = \text{Min} \{ ER_2, VAL_2 \}$

  $ACR_2 = \text{Fn} \{ ER_2, ACR_2 \}$

The unknowns in the above equations are the input rate and the VC's current rate. We shall see in the next section that there are several ways of measuring VC rates and input rates, combining the feedback from the next loop, and updating the ACR of the next loop. Note that though different switches may implement different algorithms, many measure quantities such as the VC's current rate and the ABR input rate.

# 4 VS/VD Switch Design Options

In this section, we aim at answering the following questions:

- What is a VC's current rate? (4 options)
- What is the input rate? (2 options)
- Does the congestion control actions at a link affect the next loop or the previous loop? (3 options)
- When is the VC's allocation at the link (VAL) calculated? (3 options)

We will enumerate the 72 ($= 4 \times 2 \times 3 \times 3$) option combinations and then study this state space for the best combination.

## 4.1 Measuring the VC's Current Rate

There are four methods to measure the VC's current rate:

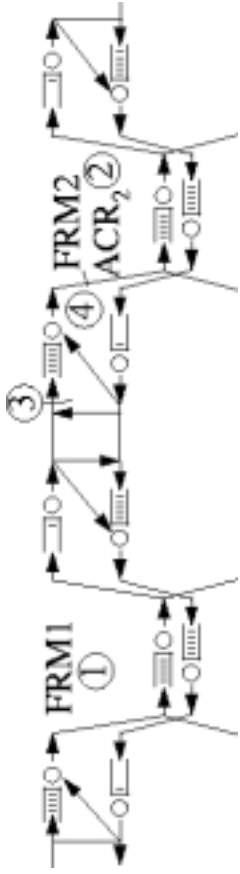

Figure 7: Four methods to measure the rate of a VC at the VS/VD switch

1. The rate of the VC is declared by the source end system of the previous loop in the Current Cell Rate (CCR) field of the FRM cell (FRM1) received by the VD. This declared value can be used as the VC's rate.

2. The VS to the next loop declares the CCR value of the FRM sent (FRM2) to be its ACR ($ACR_2$). This declared value can be used as the VC's rate.

3. The actual source rate in the *previous loop* can be measured. This rate is equal to the VC's input rate to the per-VC queue. This measured source rate can be used as the VC's rate.

4. The actual source rate in the *next loop* can be measured as the VC's input rate to the per-class queue (from the per-VC queue). This measured value can be used as the VC's rate.

Figure 7 illustrates where each method is applied (note the position of the numbers in circles).

## 4.2  Measuring the Input Rate at the Switch

Figure 8 (note the position of the numbers in circles) shows two methods of estimating the input rate for use in the switch algorithm calculations. These two methods are:

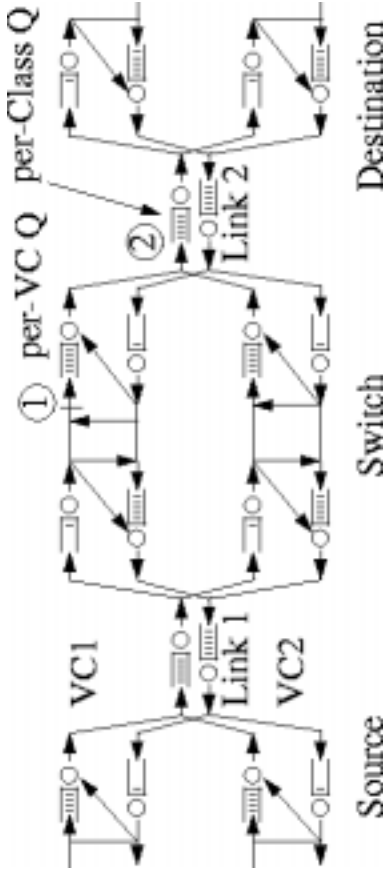

Figure 8: Two methods to measure the input rate at the VS/VD switch

1. The input rate is the sum of input rates to the per-VC ABR queues.

2. The input rate is the aggregate input rate to the per-class ABR queue.

## 4.3  Effect of Link Congestion Actions on Neighboring Links

The link congestion control actions can affect neighboring links. The following actions are possible in response to the link congestion of Link2:

1. Change $ER_1$. This affects the rate of the *previous loop only*. The change in rate is experienced only after a feedback delay equal to twice the propogation delay of the loop.

2. Change $ACR_2$. This affects the rate of the *next loop only*. The change in rate is experienced instantaneously.

3. Change $ER_1$ and $ACR_2$. This affects *both the previous and the next loop*. The next loop is affected instantaneously while the previous loop is affected after a feedback delay as in the first case.

## 4.4 Frequency of Updating the Allocated Rate

The ERICA algorithm in a non-VS/VD switch calculates the allocated rate when a BRM cell is processed in a switch. However, in a VS/VD switch, there are three options as shown in Figure 9:

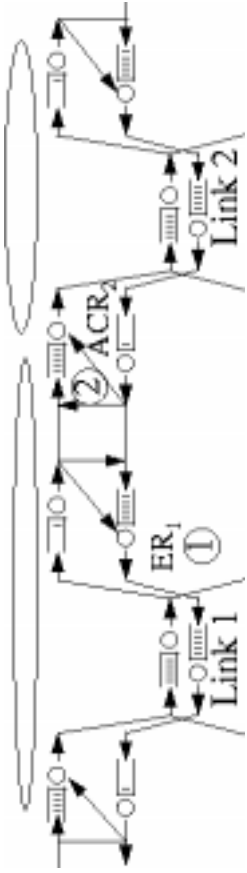

Figure 9: Three methods to update the allocated rate

1. Calculate allocated rate *on receiving BRM2 only*. Store the value in a table and use this table value when an FRM is turned around.

2. Calculate allocated rate *only when FRM1 is turned around*.

3. Calculate allocated rate *both when FRM1 is turned around as well as when BRM2 is received*.

In the next section, we discuss the various options and present analytical arguments to eliminate certain design combinations.

# 5 VS/VD Switch Design Options

## 5.1 VC Rate Measurement Techniques

We have presented four ways of finding the the VC's current rate in section 4.1, two of them used declared rates and two of them measured the actual source rate. We show that measuring source rates is better than using declared rates for two reasons.

First, the declared VC rate of a loop naively is the minimum of bottleneck rates of *downstream loops only*. It does not consider the bottleneck rates of upstream loops, and may or may not consider the bottleneck rate of the first link of the next loop. Measurement allows better estimation of load when the traffic is not regular.

Second, the actual rate of the VC may be lower than the declared ACR of the VC due to dynamic changes in bottleneck rates upstream of the current switch. The difference in ACR and VC rate will remain *at least* as long as the time required for new feedback from the bottleneck in the path to reach the source plus the time for the new VC rate to be experienced at the switch. The sum of these two delay components is called the "feedback delay." Due to feedback delay, it is possible that the declared rate is a stale value at any point of time. This is especially true in VS/VD switches where per-VC queues may control source rates to values quite different from their declared rates.

Further, the measured source rate can easily be calculated in a VS/VD switch since the necessary quantities (number of cells and time period) are measured as part of one of the source end system rules (SES Rule 5) [1, 2, 10].

## 5.2 Input Rate measurement techniques

As discussed earlier, the input rate can be measured as the sum of the input rates of VCs to the per-VC queues or the aggregate input rate to the per-class queue. These two rates can be different because the input rate to the per-VC queues is at the previous loop's rate while the input to the per-class queue is related to the next loop's rate. Figure 10 shows a simple case where two adjacent loops can run at very different rates (10 Mbps and 100Mbps) for one feedback delay.

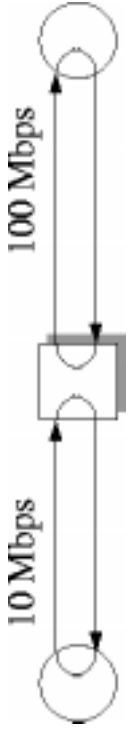

Figure 10: Two adjacent loops may operate at very different rates for one feedback delay

## 5.3 Combinations of VC rate and input rate measurement options

Table 1 summarizes the option combinations considering the fact that two adjacent loops may run at different rates. The table shows that four of these combinations may work satisfactorily. The other combinations use inconsistent information and hence may either overallocate rates leading to unconstrained queues or result in unnecessary oscillations. We can eliminate some more cases as discussed below.

Table 1: Viable combinations of VC rate and input rate measurement

| #   | VC Rate Method    | Σ VC rates (Mbps) | Input Rate Method | Input Rate Value | Design (YES/NO) |
|-----|-------------------|-------------------|-------------------|------------------|-----------------|
| 1.  | From FRM1         | 10                | Σ per-VC          | 10               | **YES**         |
| 2.  | From FRM1         | 10                | per-class         | 10-100           | NO              |
| 3.  | From FRM2         | 100               | Σ per-VC          | 10               | NO              |
| 4.  | From FRM2         | 100               | per-class         | 100              | **YES**         |
| 5.  | At per-VC queue   | 10                | Σ per-VC          | 10               | **YES**         |
| 6.  | At per-VC queue   | 10                | per-class         | 10-100           | NO              |
| 7.  | At per-class queue| 100               | Σ per-VC          | 10               | NO              |
| 8.  | At per-class queue| 100               | per-class         | 100              | **YES**         |

The above table does not make any assumptions about the queue lengths at any of the queues (per-VC or per-class). For example, when the queue lengths are close to zero, the actual source rate might be much lower than the declared rate in the FRMs leading to overallocation of rates. This criterion can be used to reject more options.

The performance of one such rejected case is shown in Figure 11 (corresponding to row 4 in Table 1). The configuration used has two ABR infinite sources and one high priority VBR source contending for the bottleneck link's (LINK1) bandwidth. The VBR has an ON/OFF pattern, where it uses 80% of the link capacity when ON. The ON time and the OFF time are equal (20 ms each). The VS/VD switch overallocates rates when the VBR source is OFF. This leads to ABR queue backlogs when the VBR source comes ON in the next cycle. The queue backlogs are never cleared, and hence the queues diverge. *In this case, the fast response of VS/VD is harmful* because the rates are overallocated.

In this study, we have not evaluated row 5 of the table (measurement of VC rate at entry to the per-VC queues). Hence, out of the total of 8 combinations, we consider two viable combinations: row 1 and row 8 of the table. Note that since row 8 uses source rate measurement, we expect it to show better performance. This is substantiated by our simulation results presented later in the paper.

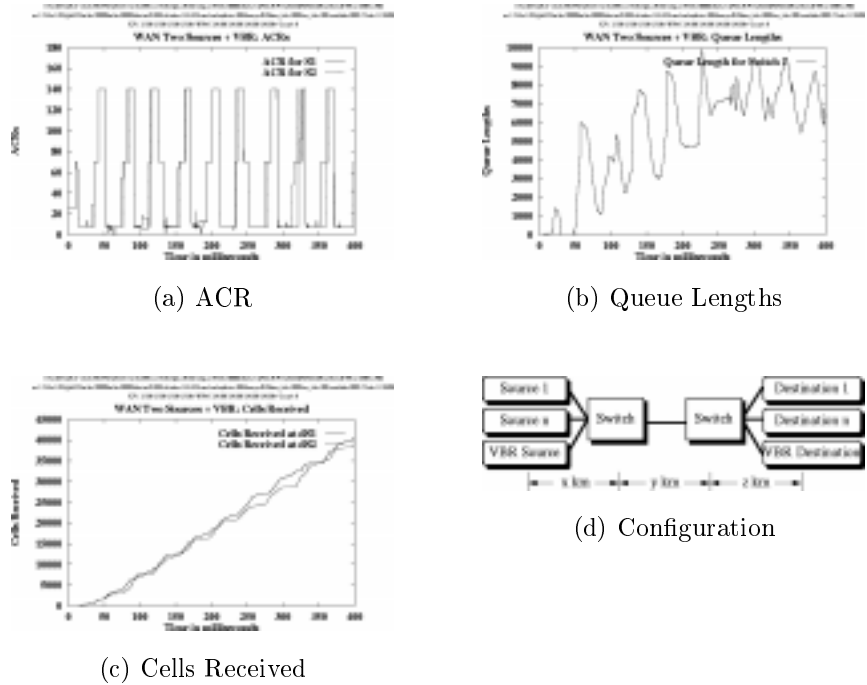

Figure 11: 2-source+VBR configuration. Unconstrained queues due to overallocation.

## 5.4 Effect of Link Congestion Control Actions

In a network with non-VS/VD switches only, the bottleneck rate needs to reach the sources before any corresponding load change is seen in the network. However, a VS/VD switch can enforce the new bottleneck rate immediately (by changing the ACR of the per-VC queue at the VS). This rate enforcement affects the utilization of links in the next loop. Hence, the VS/VD link congestion control actions can affect neighboring loops. We have enumerated three options in an earlier section.

We note that the second option ("next loop only") does not work because the congestion information is not propagated to the sources of the congestion (as required by the standard [1]). This leaves us with two alternatives. The third option ("both loops") is attractive because, when $ACR_2$ is updated, the switches in the next loop experience the load change faster. Switch algorithms may save a few iterations and converge faster in these cases.

Figure 12 shows the fast convergence in a parking lot configuration when such a VS/VD switch is used. The parking lot configuration (Figure 12(c))consists of three VCs contending for the Sw2-Sw3 link bandwidth. Link lengths are 1000 km and link bandwidths are 155.52 Mbps. The target rate of the ERICA algorithm was 90% of link bandwidth i.e., 139.97 Mbps. The round trip time for the S3-D3 VC is shorter than the round trip time for the other two VCs. The optimum allocation by ERICA for each source is 1/3 of the target rate on the Sw2-Sw3 (about 46.7 Mbps). Figure 12(a) shows that the optimum value is reached at 40 ms. Part (b) of the figure shows that the transient queues are small and that the allocation is fair.

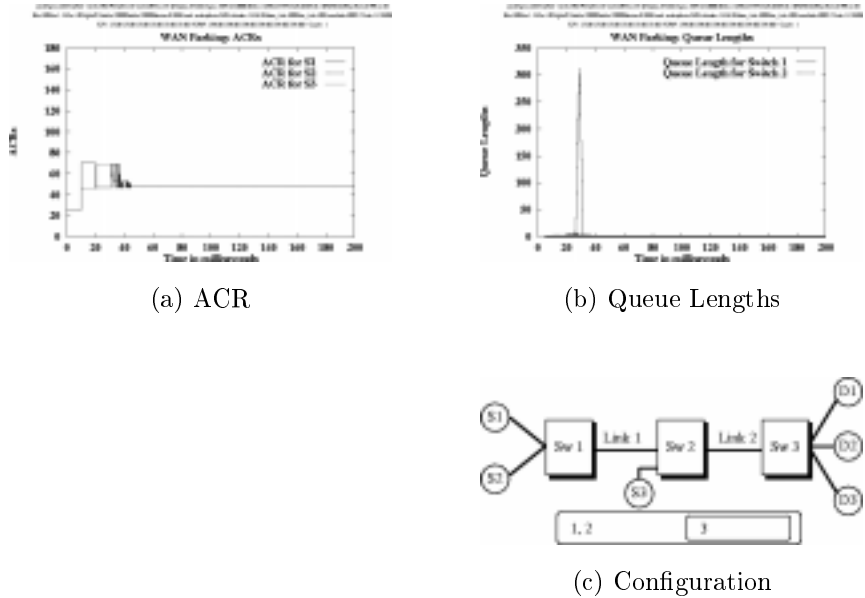

(a) ACR  (b) Queue Lengths

(c) Configuration

Figure 12: Parking lot configuration. Illustrates fast convergence of the best VS/VD option.

fig:example-conv

## 5.5 Link Congestion and Allocated Rate Update Frequency: Viable Options

The allocated rate update has three options:
a) update upon BRM receipt (in VS) and enter the value in a table to be used when an FRM is turned around,
b) update upon FRM turnaround (at VD) and no action at VS,
c) update both at FRM (VD) and at BRM (VS) without use of a table.

The last option recomputes the allocated rate a larger number of times, but can potentially allocate rates better because we always use the latest information.

The allocated rate update and the effects of link congestion actions interact as shown in Figure 13. The figure shows a tree where the first level considers the link congestion (2 options), i.e., whether the next loop is also affected or not. The second level lists the three options for the allocated rate update frequency. The viable options are those highlighted in bold at the leaf level.

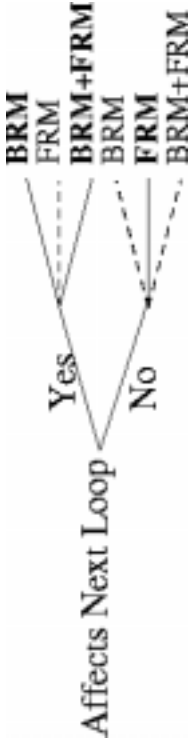

Figure 13: Link congestion and allocated rate update: viable options

Other options are not viable because of the following reasons. In particular, if the link congestion does not affect the next loop, the allocated rate update at the FRM turnaround is all that is required. The allocated rate at the BRM is redundant in this case. Further, if the link congestion affects the next loop, then the allocated rate update has to be done on receiving a BRM, so that ACR can be changed at the VS. This gives us two possibilities as shown in the figure (BRM only, and BRM+FRM).

Hence, we have three viable combinations of link congestion and the allocated rate update frequency. A summary of all viable options (a total of 6) is listed in Table 2.

Table 2: Summary of viable VS/VD design alternatives

| Option # | VC Rate Method | Input Rate Measurement point | Link Congestion Effect | Allocated Rate Updated at |
|---|---|---|---|---|
| 41 | From FRM1 | per-VC | previous loop only | FRM1 only |
| 52 | measured at per-class Q | per-class | both loops | FRM1 only |
| 329 | From FRM1 | per-VC | both loops | FRM1 only |
| 340 | measured at per-class Q | per-class | both loops | FRM1 and BRM2 |
| 393 | From FRM1 | per-VC | both loops | BRM2 only |
| 404 | measured at per-class Q | per-class | both loops | BRM2 only |

The next section evaluates the performance of the viable VS/VD design options through simulation.

# 6 Performance Evaluation of VS/VD Design Options

## 6.1 Metrics

We use four metrics to evaluate the performance of these alternatives:

- **Response Time:** is the time taken to reach near optimal behavior on startup.
- **Convergence Time:** is the time for rate oscillations to decrease (time to reach the steady state).
- **Throughput:** Total data transferred per unit time.
- **Maximum Queue:** The maximum queue before convergence.

The difference between response time and convergence time is illustrated in Figure 14. The following sections present simulation results with respect to the above metrics. Note that we have used greedy (infinite) traffic sources in our simulations. We have studied the algorithmic enhancements in non-VS/VD switches for non-greedy sources in reference [8]. We expect consistent results for such traffic when the best implementation option (see below) is used.

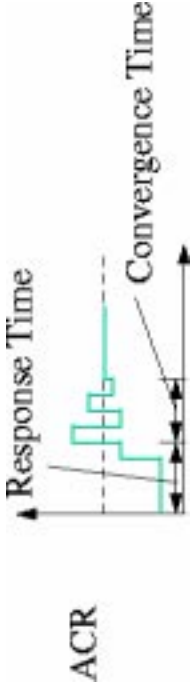

Figure 14: Response time vs Convergence time

## 6.1.1 Response Time

Without VS/VD all response times are close to the round-trip delay. With VS/VD, the response times are close to the feedback delay from the bottleneck. Since VS/VD reduces the response time during the first round trip, it is good for long delay paths. The quick response time (10 ms in the parking lot configuration which has a 30 ms round trip time) was illustrated previously in Figure 12. Response time is also important for bursty traffic like TCP file transfer over ATM which "starts up" at the beginning of every active period (when the TCP window increases) after the corresponding idle period [9, 10].

### 6.1.2 Throughput

The number of cells received at the destination is a measure of the throughput achieved. These values are listed in Table 3. The top row is a list of the configuration codes (these codes are explained in Table 2. The final column lists the throughput values for the case when a non-VS/VD switch is used. The 2 source+VBR and the parking lot configurations have been introduced in earlier section.

The upstream bottleneck configuration shown in Figure 15 has a bottleneck at Sw1 where 15 VCs share the Sw1-Sw2 link. As a result the S15-D15 VC is not capable of utilizing its bandwidth share at the Sw2-Sw3 link. This excess bandwidth needs to be shared equally by the other two VCs. The table entry shows the number of cells received at the destination for either the S16-D16 VC or the S17-D17 VC.

In the 2 source+VBR and the upstream bottleneck configurations, the simulation was run for 400 ms (the destination receives data from time = 15 ms through 400 ms). In the parking lot configuration, the simulation was run for 200ms.

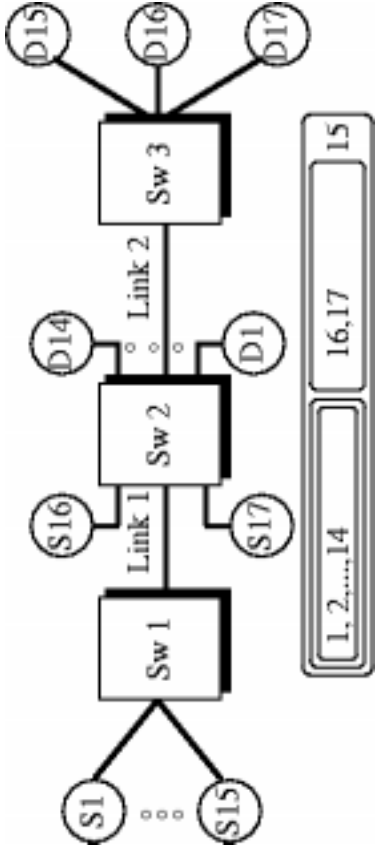

Figure 15: Upstream bottleneck configuration

Table 3: Cells received at the destination per source in Kcells

| VS/VD Option # → | A | B | C | D | E | F | No VS/VD |
|---|---|---|---|---|---|---|---|
| Configuration ↓ | | | | | | | |
| 2 source + VBR | 31 | 31 | 32.5 | 34 | 32 | 33 | 30 |
| Parking lot | 22 | 22 | 23 | 20.5 | 23 | 20.5 | 19.5 |
| Upstream bottleneck | 61 | 61 | 61 | 60 | 61 | 61 | 62 |

As we compare the values in each row of the table, we find that, in general, there is *little difference between the alternatives in terms of throughput.* However, there is a slight increase in throughput when VS/VD is used over the case without VS/VD switch.

### 6.1.3 Convergence Time

The convergence time is a measure of how fast the scheme finishes the transient phase and reaches steady state. It is also sometimes called "transient response." The convergence times of the various options are shown in Table 4. The "transient" configuration mentioned in the table has two ABR VCs sharing a bottleneck (like the 2 source + VBR configuration, but without the VBR VC). One of the VCs comes on in the middle of the simulation and remains active for a period of 60 ms before going off.

Table 4: Convergence time in ms

| VS/VD Option # → | A | B | C | D | E | F | No VS/VD |
|---:|---|---|---|---|---|---|---|
| Configuration ↓ | | | | | | | |
| Transient | 50 | 50 | 65 | **20** | 55 | 25 | 60 |
| Parking lot | 120 | 100 | 170 | **45** | 125 | 50 | 140 |
| Upstream bottleneck | 95 | 75 | 75 | **20** | 95 | **20** | 70 |

Observe that the convergence time of VS/VD option D (highlighted) is the best. Recall that this configuration corresponds to measuring the VC rate at the entry to the per-class queue, input rate measured at the per-class queue, link congestion affecting both the next loop and the previous loop, the allocated rate updated at both FRM1 and BRM2.

### 6.1.4 Maximum Transient Queue Length

The maximum transient queues gives a measure of how askew the allocations were when compared to the optimal allocation and how soon this was corrected. The maximum transient queues are tabulated for various configurations for each VS/VD option and for the case without VS/VD in Table 5.

The table shows that VS/VD option D has very small transient queues in all the configurations and the minimum queues in a majority of cases. This result, combined with the fastest response and near-maximum throughput behavior confirms the choice of option D as the best VS/VD implementation.

Table 5: Maximum queue length in Kcells

| VS/VD Option # → | A | B | C | D | E | F | No VS/VD |
|---|---|---|---|---|---|---|---|
| Configuration ↓ | | | | | | | |
| 2 Source + VBR | **1.2** | 1.4 | 2.7 | 1.8 | 2.7 | 1.8 | 2.7 |
| Transient | 1.4 | 1.1 | 1.4 | **0.025** | 1.3 | 1.0 | 6.0 |
| Parking lot | 1.9 | 1.9 | 1.4 | **0.3** | 3.7 | 0.35 | 2.0 |
| Upstream bottleneck | 0.025 | 0.08 | 0.3 | **0.005** | 1.3 | **0.005** | 0.19 |

Observe that the queues for the VS/VD implementations are in general lesser than or equal to the queues for the case without VS/VD. However, the queues reduce much more if the correct implementation (like option D) is chosen.

# 7 Conclusions

In summary:

- VS/VD is an option that can be added to switches which implement per-VC queueing. The addition can potentially yield improved performance in terms of response time, convergence time, and smaller queues. This is especially useful for switches at the edge of satellite networks or switches that are attached to links with large delay-bandwidth product. The fast response and convergence times also help support bursty traffic like data more efficiently.

- The effect of VS/VD depends upon the switch algorithm used and how it is implemented in the VS/VD switch. The convergence time and transient queues can be very different for different VS/VD implementations of the same basic switch algorithm. In such cases the fast response of VS/VD is harmful.

- With VS/VD, ACR and actual rates are very different. The switch cannot rely on the RM cell CCR field. We recommend that the VS/VD switch and in general, switches implementing per-VC queueing measure the VC's current rate.

- The sum of the input rates to per-VC VS queues is not the same as the input rate to the link. It is best to measure the VC's rate at the output of the VS and the input rate at the entry to the per-class queue.

- On detecting link congestion, the congestion information should be forwarded to the previous loop as well as the next loop. This method reduces the convergence time by reducing the number of iterations required in the switch algorithms on the current and downstream switches.

- It is best for the the rate allocated to a VC to be calculated both when turning around FRMs at the VD as well as after receiving BRMs at the next VS.

# 8 Future Work

The VS/VD provision in the ABR traffic management framework can potentially improve performance of bursty traffic and reduce the buffer requirements in switches. The VS/VD mechanism achieves this by breaking up a large ABR loop into smaller ABR loops which are separately controlled. However, further study is required in the following areas:

- Effect of VS/VD on buffer requirements in the switch.
- Scheduling issues with VS/VD.
- Effect of different switch algorithms in different control loops, and different control loop lengths.
- Effect of non-ABR clouds and standardization issues involved.
- Effect of using switch algorithms specifically designed to exploit the per-VC queueing policy required in VS/VD implementations.

---

[2] All our papers and ATM Forum contributions are available through http://www.cis.ohio-state.edu/~jain